\documentclass[prc,twocolumn,secnumroman,tightenlines,amssymb,aps,nobibnotes,
  superscriptaddress,showpacs,balancelastpage,floatfix,nofootinbib]{revtex4}
\usepackage{graphics}
\usepackage{graphicx,colordvi}
\usepackage{multirow,color}
\usepackage{color}
\usepackage[utf8]{inputenc}
\usepackage{amsmath}

\begin{document}

\title{Search for shape-isomers in the Pt-Hg-Pb region}


\author{J. Bartel}\email{Johann.Bartel@iphc.cnrs.fr}
\affiliation{D.R.S. Institut Pluridisciplinaire Hubert Curien, CNRS-IN2P3, Universit\'e de Strasbourg, Strasbourg, France}
\author{H. Molique}
\affiliation{D.R.S. Institut Pluridisciplinaire Hubert Curien, CNRS-IN2P3, Universit\'e de Strasbourg, Strasbourg, France}
\author{B. Nerlo-Pomorska}
\affiliation{Department of Theoretical Physics, Maria Curie-Sk\l odowska University, Lublin, Poland}
\author{M. Warda}
\affiliation{Department of Theoretical Physics, Maria Curie-Sk\l odowska University, Lublin, Poland}
\author{K. Pomorski}
\affiliation{National Centre for Nuclear Research, Warsaw, Poland}

\pacs{21.10.Dr, 21.10.Ma, 21.60.Cs, 21.60.Jz, 25.85.Ca}
\date{\today}

\begin{abstract}
\noindent

Potential energy surfaces of nine even-even isotopes of Pt, Hg, and Pb around $^{186}$Pt are evaluated within a macroscopic-microscopic model based on the Lublin-Strasbourg-Drop macroscopic energy and the microscopic energy obtained using the Yukawa-folded mean-field potential to establish the Strutinski shell corrections and the 
pairing correlation energy through the BCS approach with a monopole pairing force. The rapidly converging Fourier-over-Spheroid shape parametrization is used to describe nuclear deformations. The stability of the identified shape isomeric states with respect to non-axial and higher-order deformations is investigated. It is also found that in the description of non-axial deformations special attention needs to be devoted to the orientation of the triaxial shape.
For the example of the $^{186}$Hg nucleus, where three prolate shape-isomeric states are found, it is shown that the potential energy surface obtained in our model is  close to the one obtained in the Hartree-Fock-Bogoliubov theory with the Gogny energy-density fuctional.\\

\noindent
KEYWORDS: macro-micro model, shape coexistence, shape isomers.

\end{abstract}

\maketitle 
 

\section{Introduction} 
\label{intro} 

Our understanding of shape isomerism and shape coexistence in nuclei has substantially changed over the last decades. Even though shape isomers were discovered in the 1960s in connection with investigations of fissioning systems \cite{Po62}, their existence was long understood as an exotic phenomenon limited to different islands in the nuclear chart. The present understanding, however, is the one of the occurrence of such states in nuclei all along the nuclear chart, with the only exception of very light nuclei \cite{WH16,GZ22,YZ23,BM23a,BM23b,WJ24}. The existence in $^{66}$Ni, to give just one prominent example, of (prolate as well as oblate) isomeric states has been identified both through the experiment (see, e.g., Ref.\ \cite{LF17,LF24} and references therein) and a large variety of very different theoretical models \cite{BK89,GD89} ranging from mean-field calculations of the Hartree-Fock + BCS or Hartree-Fock-Bogoliubov type to macroscopic-microscopic approaches \cite{MSB09,MSB12}, to Monte-Carlo Shell-Model calculations of the Takaharu Otsuka group \cite{LVF23} or Shell Model calculations with LNPS interaction of the Madrid-Strasbourg group \cite{OFM17}.

The region just below the $Z=82$ shell closure (Platinum, Mercury, and Lead) has already been identified some time ago in different studies  \cite{CW90,AHD00,HWo11,YBH13,Po16,NP17} as one of the most favorable for the occurrence of shape coexistence. This discovery has strongly motivated theoretical studies to explore this region of nuclei in more detail. Important advances in the study of neutron-deficient nuclei around $Z\approx 82$ have been realized using shape isomers and $\beta$-decays of nuclei populated by relativistic energy-fragmentation experiments \cite{Pod12} performed at GSI within the RISING campaign or through tagging techniques at the Accelerator Laboratory of the University of Jyväskylä (Finland) \cite{JGP16}, and through Coulomb-excitation experiments undertaken at the REX-ISOLDE facility in CERN~\cite{WLG16}.

Interesting theoretical studies on this subject, based either on a self-consistent approach (see, e.g., Refs.\ \cite{BHR03,YBH13,NRR13,NVR02}) or on macroscopic-microscopic (e.g., \cite{MSB09,MSB12,PND20}) models, have been performed in the past.

The aim of the present paper is to study the influence on the potential energy surfaces (PES) of even-even Pt, Hg, and Pb isotopes around $^{186}$Hg of higher-order deformations that go beyond a simple quadrupole shape using our recently developed and very efficient {\it Fourier-over-Spheroid} (FoS) shape parametrization \cite{PNP23,PNB23}, together with the macroscopic-microscopic (mac-mic) approach with the Lublin-Strasbourg Drop (LSD) \cite{PDu03}, able to reproduce accurately both nuclear masses and fission barriers, and a Yukawa-folded single-particle potential \cite{DNi76,DPB16} to account for shell and pairing-energy corrections. The results will be compared with the outcome of the self-consistent calculations performed in the Hatree-Fock-Bogoliubov (HFB) theory with Gogny D1S force \cite{BGG84} .

Although the investigations of the nuclear potential-energy surfaces presented in a previous study \cite{PND20} had been carried out for a broader range of nuclei in a similar approach, some new effects are included in the present work through our new FoS shape parametrization which enables us, in particular, to project the evaluated PES onto the traditional $(\beta,\gamma)$  plane. 
Let us recall that the potential-energy surfaces in Ref.\ \cite{PND20} had been evaluated in the space of the original Fourier deformation parameters $q_2,q_3,q_4,\eta$, (with $\eta$ standing for the non-axiality degree of freedom). One of the aims of our present study, in addition of being more precise due to the use of our new ``{\it Fourier-over-Spheroid}'' shape parametrization and a denser deformation mesh, is, however, to show our potential-energy surfaces in terms of the much more widely used Bohr shape parameters $\{\beta,\gamma\}$, 
what enables us to take into account effect of different nuclear orientations in space and evaluate more precisely the effect of higher-order deformations. 
In addition, higher-order deformation parameters are taken into account here, still keeping the 60$^{\circ}$ symmetry of the original $(\beta,\gamma)$  parametrization.

\section{Shapes breaking axial symmetry}

A search of possible shape isomers is carried out in the Pt, Hg, Pb (Z=78, 80, 82) region by studying the landscapes in the deformation-energy space of different even-even isotopes of each of these nuclei. The calculation of the nuclear 
energy as a function of the chosen deformation parameters is performed in the framework of the macroscopic-microscopic approach with the LSD model \cite{PDu03} for the macroscopic term and the shell and pairing energy corrections evaluated using the Yukawa-folded mean-field potential \cite{DNi76} for the microscopic part.
To be able to have at our disposal a most flexible description of the nuclear deformation, we use the Fourier shape  parametrization \cite{SPN17} in the form of what we call the ``Fourier over Spheroid'' definition \cite{PNP23,PNB23}
with a deformation space spanned by four parameters $\left\{c,\eta,a_3,a_4\right\}$ which stand respectively for elongation, non-axiality, left-right asymmetry, and neck (hexadecapole type) degree of freedom. 

Our aim is to show that this deformation space is in a way equivalent to the widely known and used Bohr $\left\{\beta,\gamma\right\}$ shape parametrization \cite{Boh52}, 
though allowing for substantially more general deformations. 
An ellipsoid can, indeed, be generally described in cartesian 
coordinates by three half-axis ${\cal A}_i$, ${\cal A}_j$ and ${\cal A}_k$ through the equation
\begin{equation}
{x^2\over{\cal A}_1^2} + {y^2\over{\cal A}_2^2} +{z^2\over{\cal A}_3^2}  =1~.
\end{equation}
If all the half-axis are different, one speaks about a Jacobi spheroid (symmetry group $D_{2h}$). If two of the half-axis are the same, like 
${\cal A}_1={\cal A}_2$, the spheroid has a symmetry axis. One then speaks about a spheroid of revolution, which can be prolate or oblate (Maclaurin spheroid). Instead of the half-axis ${\cal A}_\nu$, one then introduces 
deformation parameters $\beta$ and $\gamma$ through
\begin{equation}
 {\cal A}_\nu = R(\beta,\gamma) \left[1 + k\,\cos(\gamma - \frac{2\pi}{3} \nu) \right]
  \, , \;\;\; \nu = 1,2,3
\end{equation}
where $k=\sqrt{\frac{5}{4\pi}} \beta$ and where 
\begin{equation}
  R(\beta,\gamma) = R_0 \left[1 - \frac{3}{4}k^2 + \frac{k^3}{4} \cos(3 \gamma) \right]^{-1/3}
\end{equation}
with $R_0$ being the radius of the spherical shape and where volume conservation 
requires that $\frac{4\pi}{3} {\cal A}_1 {\cal A}_2 {\cal A}_3 = \frac{4\pi}{3} R_0^3$. 

\begin{figure}[!ht]
\centering
\includegraphics[width=\columnwidth]{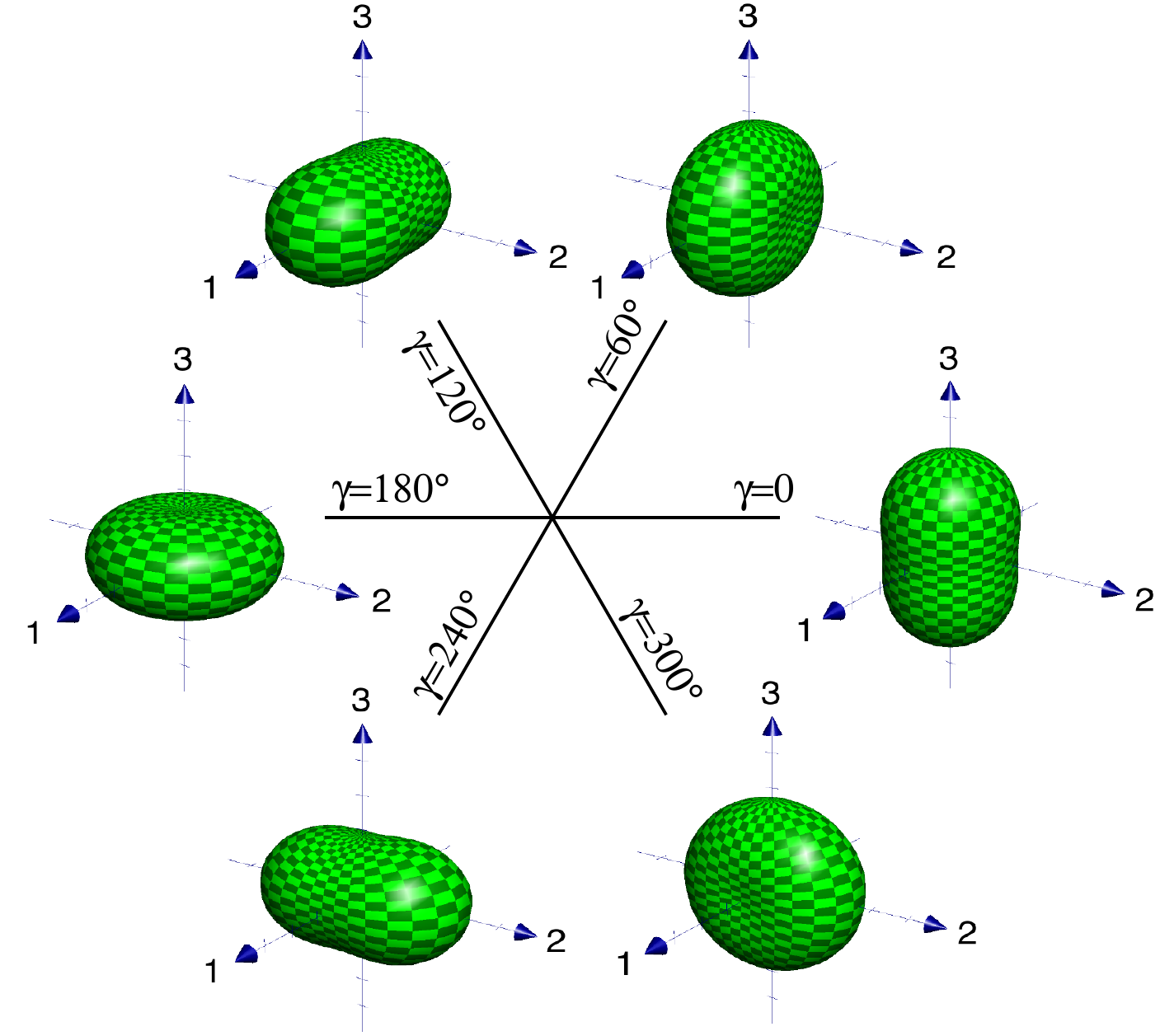} \\[ 0.2cm]
\caption{Illustration of the symmetry of Bohr's $\{\beta,\gamma\}$ shape parametrization.}
\label{Bohrbg}
\end{figure}

The Bohr parametrization allows to describe non-axial spheroidal shapes with 
two deformation parameters, $\beta$ for its elongation and $\gamma$, an angle that takes care of the orientation of the spheroid, as illustrated in Fig. \ref{Bohrbg}. The above equations show that $\beta = 0$ corresponds to the spherical shape. Along the abscissa ($\gamma \!=\! 0^\circ$), one has a prolate deformation with $z$ (or 3) as the symmetry axis, while for $\gamma = 60^\circ$ the shape is oblate with $y$ (or 2) as the symmetry axis. The $\{\beta,\gamma\}$ plane is then symmetric in $\gamma = 60^\circ$ portions so that the shape with $\gamma = 120^\circ$ is identical to the one with $\gamma = 0^\circ$ except that the profile is now symmetric through rotations about the $x$ (or 1) axis which is the symmetry axis. An identical nuclear deformation appears 3 times, each time at an interval of 120 degrees with a symmetry axis, which is respectively the 1, 2 or 3-axis.

If higher order, such as hexadecapole deformations, need to be included, 
this $60^\circ$ symmetry is broken, and a given $\{\beta,\gamma\}$ deformation 
is no longer the same (not even with a different orientation) as the 
corresponding $\{\beta,\gamma +120^\circ\}$, or $\{\beta,\gamma +240^\circ\}$ deformation. This is, however, the difficulty that one encounters when trying to represent a nuclear shape obtained in the Fourier shape parametrization \cite{SPN17,PNP23} in the framework of the Bohr $\{\beta,\gamma\}$ deformation space and, also in the Rayleigh type expansions of nuclear surfaces (confer, e.g., \ Ref.\ \cite{RSo81}).

Instead of limiting oneself to the $0 \!\leq\! \gamma \!\leq\! 60^\circ$ sector and exploring, for a given deformation point, higher-order deformations, one performs calculations of the potential-energy surface (PES) in all three sectors \mbox{$0 < \gamma < 60^\circ$}, \mbox{$60^\circ < \gamma < 120^\circ$}, \mbox{$120^\circ < \gamma < 180^\circ$}, which basically means that one starts from a particular nuclear shape, but oriented differently in space, and investigates which is the orientation where the addition of higher order deformations leads to the minimal energy. In this way, one restores the initial symmetry of the three sectors, allowing us to obtain what we can call the ``{\it symmetrized potential energy surface}'' (SPES), which we will discuss in the present study.

\begin{figure}[!h]
\centering
\includegraphics[width=\columnwidth]{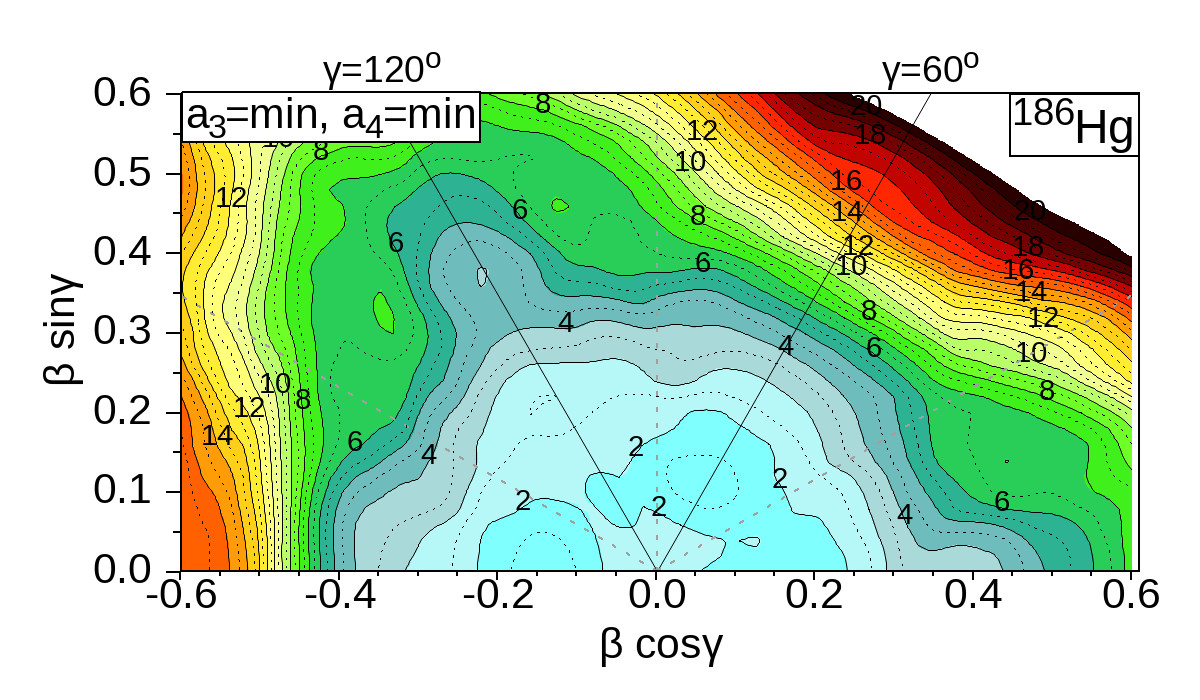}
\includegraphics[width=\columnwidth]{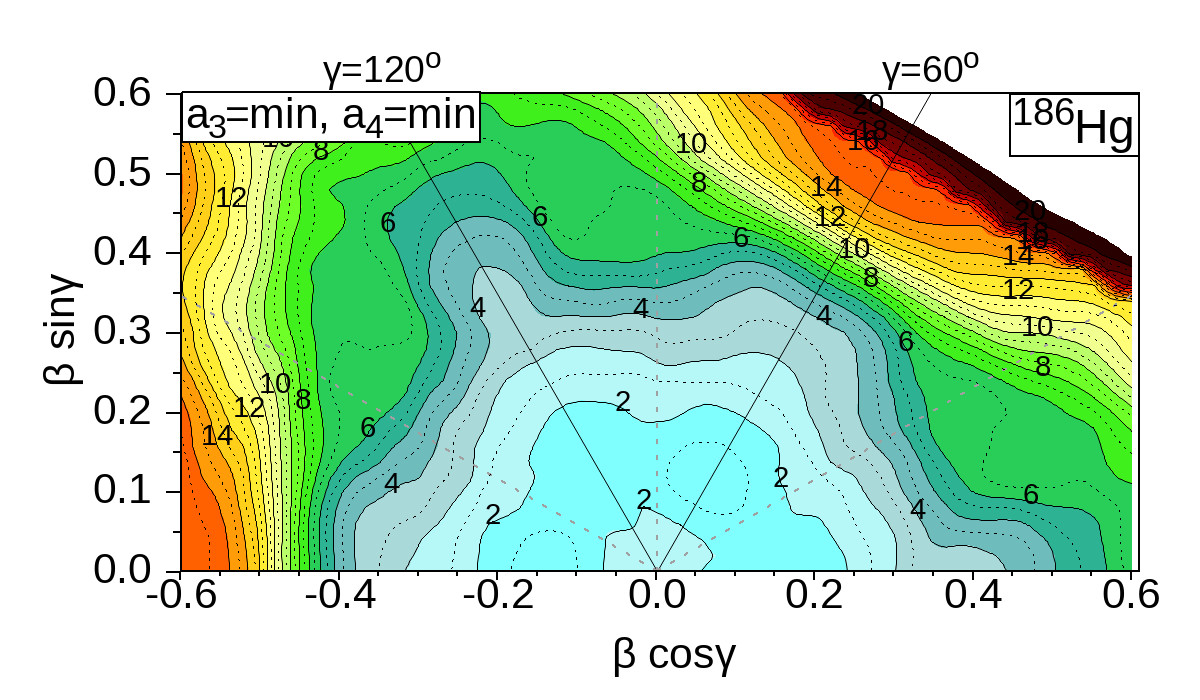}
\caption{Potential energy surface (PES) obtained in the tree $(\beta,\gamma)$ sectors (upper map) and symmetrized potential-energy surface (SPES) (lower map) for the $^{186}$Hg nucleus obtained by the symmetrization procedure as explained in the text.}
\label{186Hg_sym}
\end{figure}

In order to illustrate the method, we show in Fig.\ \ref{186Hg_sym} the direct (upper part) and the symmetrized potential energy (lower part) surfaces obtained for the $^{186}$Hg isotope, 
minimized with respect to the deformation parameters $a_3$ and $a_4$ \cite{SPN17}.
%
%
Energy labels on these and subsequent figures always indicate the total nuclear energy relative to the spherical macroscopic energy.
One notices in particular in the lower part of the figure the symmetry of the three sectors delimited by the lines $\gamma = 60^\circ$ and $\gamma = 120^\circ$ two of which then obviously become redundant.
%


\subsection{Macroscopic-microscopic model}
\label{mm}

In the mac-mic method, proposed first by Myers and \'Swi{\c a}tecki \cite{MSw66}, the total energy of the deformed nucleus is equal to the sum of a macroscopic (liquid-drop type) energy and the quantum energy correction for protons and neutrons generated by shell and pairing effects
\begin{equation}
 E_{\rm tot} = E^{}_{\rm LSD}+ {\rm E_{shell}} + {\rm E_{\rm pair}} \; .
\label{eq01}
\end{equation}
The LSD model \cite{PDu03}, which reproduces well all experimental masses and fission barrier heights, is used in this study to evaluate the macroscopic part of the energy. The shell corrections are obtained by subtracting the average energy $\widetilde E$ from the sum of the single-particle (s.p.) energies of occupied orbitals
\begin{equation}
 {\rm E_{\rm shell}} = \sum_k e_k - \widetilde E \; .
\label{eq02}
\end{equation}
As s.p.\ energies $e_k$, we have taken the eigenvalues of a mean-field Hamiltonian with the Yukawa-folded s.p. potential \cite{DNi76}. The average energy $\widetilde E$ is evaluated using the Strutinsky prescription \cite{Str66, NTS69} with a 6$^{\rm th}$ order correction polynomial. The pairing energy correction is determined as the difference between the BCS energy \cite{BCS57} and the s.p. energy sum from which the average pairing energy \cite{NTS69} is subtracted
\begin{equation}
 E_{\rm pair} = E_{\rm BCS} - \sum_{k} e_k
            - \widetilde{E}_{\rm pair}\;.
\label{Epair}
\end{equation}
In the BCS approximation, the ground-state energy of a system with an even number of particles is given by
\begin{equation}
E_{\rm BCS} = \sum_{k>0} 2e_k v_k^2 - G(\sum_{k>0}u_kv_k)^2 - G\sum_{k>0} v_k^4
 -{\cal E}_0^\varphi\;,
\label{EBCS}
\end{equation}
where the sums run over the pairs of s.p.\ levels belonging to the pairing window defined below. The coefficients $v_k$ and $u_k=\sqrt{1-v_k^2}$ are the BCS occupation amplitudes, and ${\cal E}_0^\varphi$ is the energy correction due to the particle number projection done in the GCM+GOA approximation \cite{GPo86}
\begin{equation}
{\cal E}_0^\varphi=\frac{\sum\limits_{k>0}[ (e_k-\lambda)(u_k^2-v_k^2)
        +2\Delta u_k v_k +Gv_k^4] / E_k^2}{\sum\limits_{k>0} E_k^{-2}}\;.
\label{Ephi}
\end{equation}
Here $E_k=\sqrt{(e_k-\lambda)^2+\Delta^2}$ are the quasi-particle energies with $\Delta$ and $\lambda$ the pairing gap and the Fermi energy, respectively. The average projected pairing energy, for a pairing window of width $2\Omega$, symmetric in energy with respect to the Fermi level, is equal to
\begin{equation}
\begin{array}{l}
 \widetilde{E}_{\rm pair} \!=\!\displaystyle{-\frac{\tilde{g}}{2}\,
 \tilde{\Delta}^2+\frac{\tilde{g}}{2}\,G\tilde{\Delta}\,
 {\rm arctan}\left(\!\frac{\Omega}{\tilde\Delta}\!\right)
  \!-\log\left(\!\frac{\Omega}{\tilde\Delta}\!\right)\tilde{\Delta}}\\[3ex]
~~~~~~~~~\displaystyle{
+\frac{3}{4}G\frac{\Omega/\tilde{\Delta}}{1+(\Omega/\tilde{\Delta})^2}/
  {\rm arctan}\left(\frac{\Omega}{\tilde{\Delta}}\right)\!-\!\frac{1}{4}G }~ ,
\end{array}
\label{Epavr}
\end{equation}
where $\tilde{g}$ is the average single-particle level density and $\tilde\Delta$ the average pairing gap corresponding to a pairing strength $G$ 
\begin{equation}
\tilde\Delta=\displaystyle{2\Omega\exp\left(-\frac{1}{G\tilde{g}}\right)}~.
\label{Davr}
\end{equation}
The pairing window for protons or neutrons contains $2\sqrt{15\cal N}$ ($\cal N =\,$N or Z) s.p. levels closest to the Fermi energy states. For such a window, the paring strength approximated in Ref.~\cite{PPS89} is given by the following expression:
\begin{equation} 
 G = \displaystyle{\frac{g_0}{{\cal N}^{2/3} \, A^{1/3}}}~.
\label{Gpair}
\end{equation}
The same value $g_0=g_0^p=g_0^n=0.28 \hbar \omega_0$ is taken for protons and neutrons, where $\hbar \omega_0 = 41$\,MeV$/A^{1/3}$ is the nuclear harmonic oscillator constant.

In our calculation, the single-particle spectra are obtained by diagonalization of the s.p. Hamiltonian with the Yukawa-folded potential \cite{DNi76, DPB16} with the same parameters as used in Ref.~\cite{MNi95}.


\section{SPES's of selected Pt, Hg, and Pb isotopes}

\begin{figure}[!ht]
\centering
\includegraphics[width=\columnwidth]{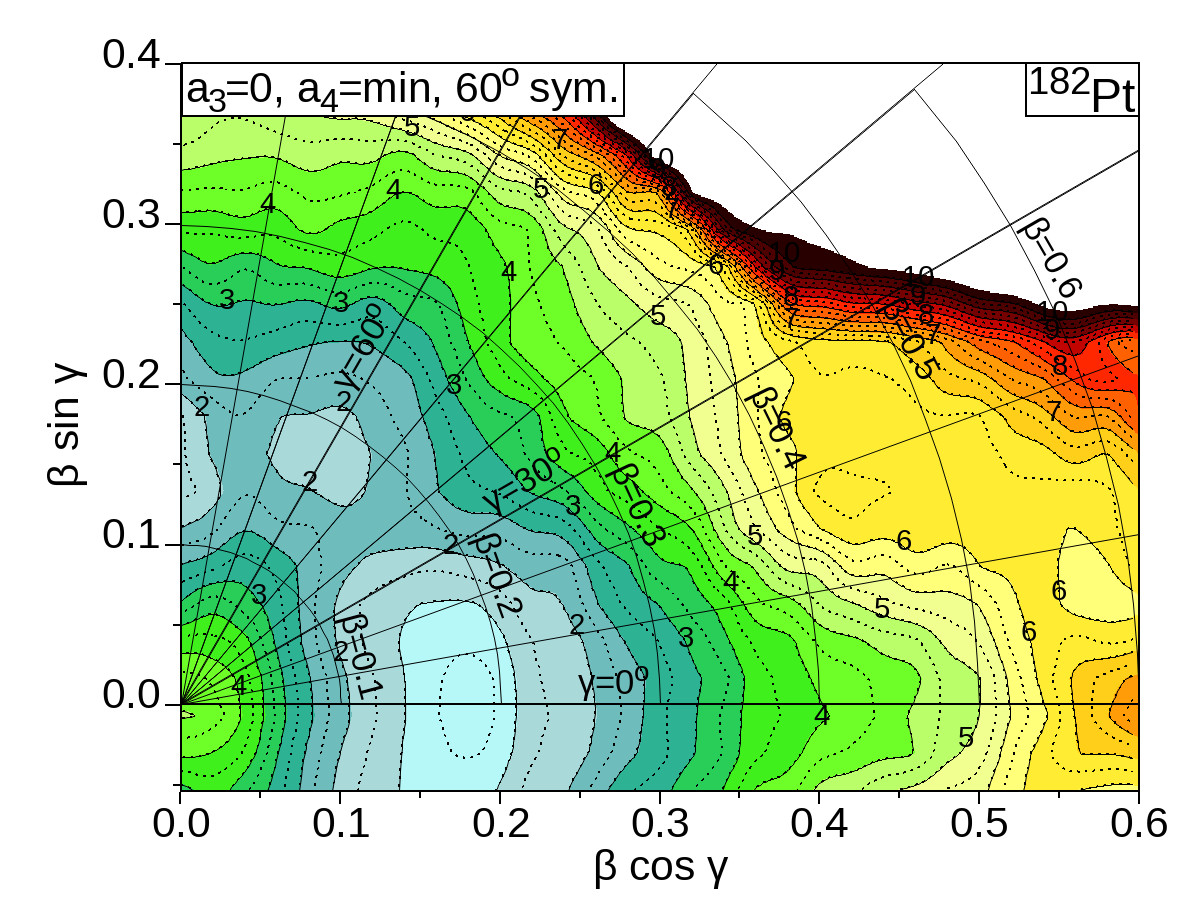} \\[0ex]
\includegraphics[width=\columnwidth]{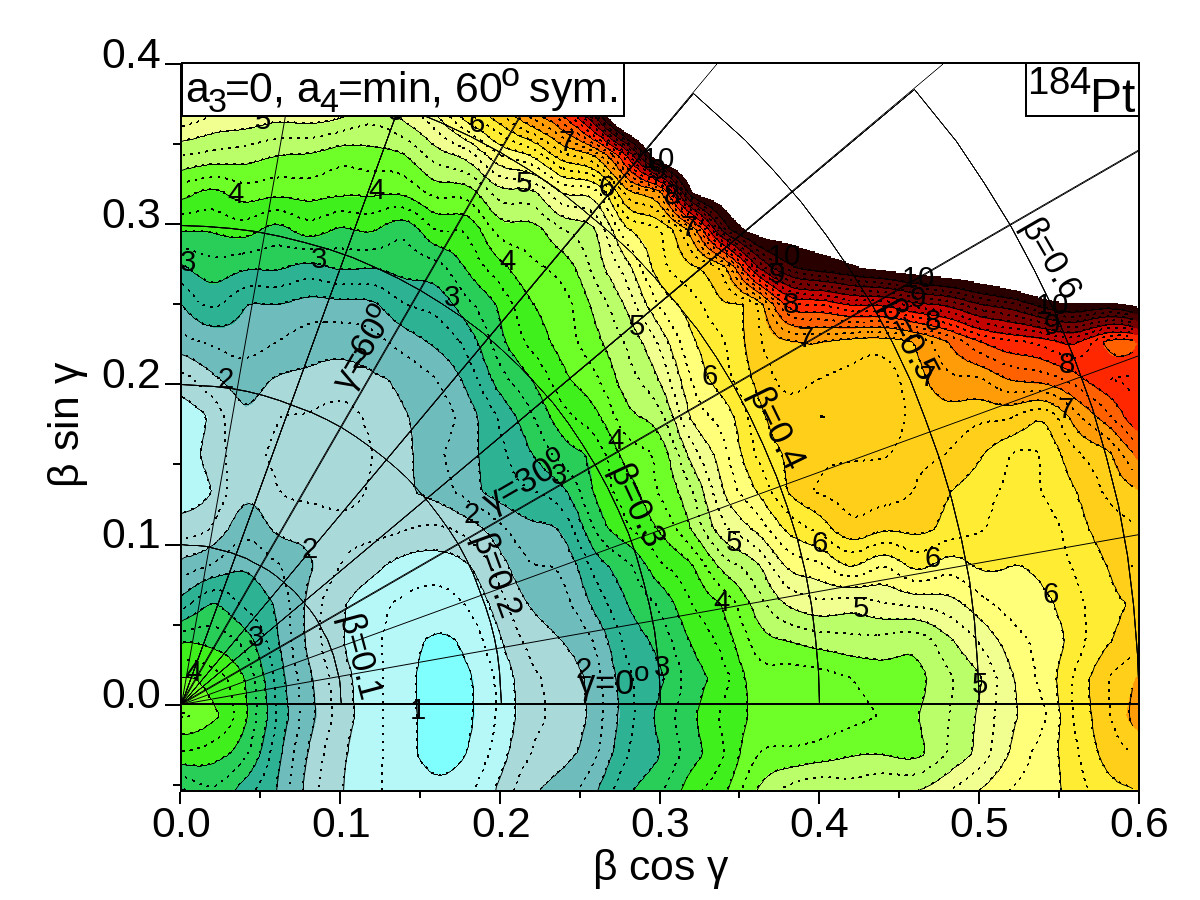} \\[0ex]
\includegraphics[width=\columnwidth]{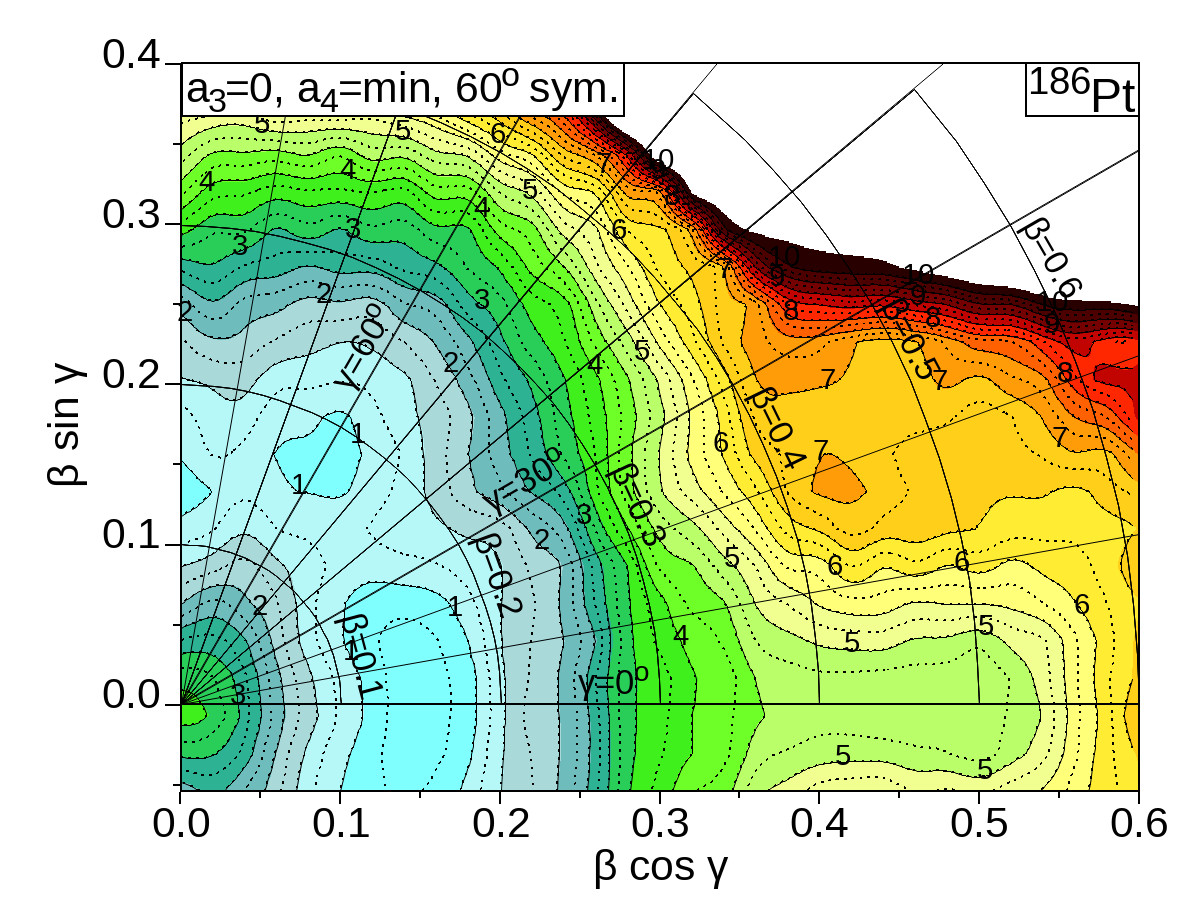} \\[0ex]
\caption{Symmetrized potential energy surfaces for Pt isotopes.}
\label{Ptmaps}
\end{figure}

\begin{figure}[!ht]
\centering
\includegraphics[width=\columnwidth]{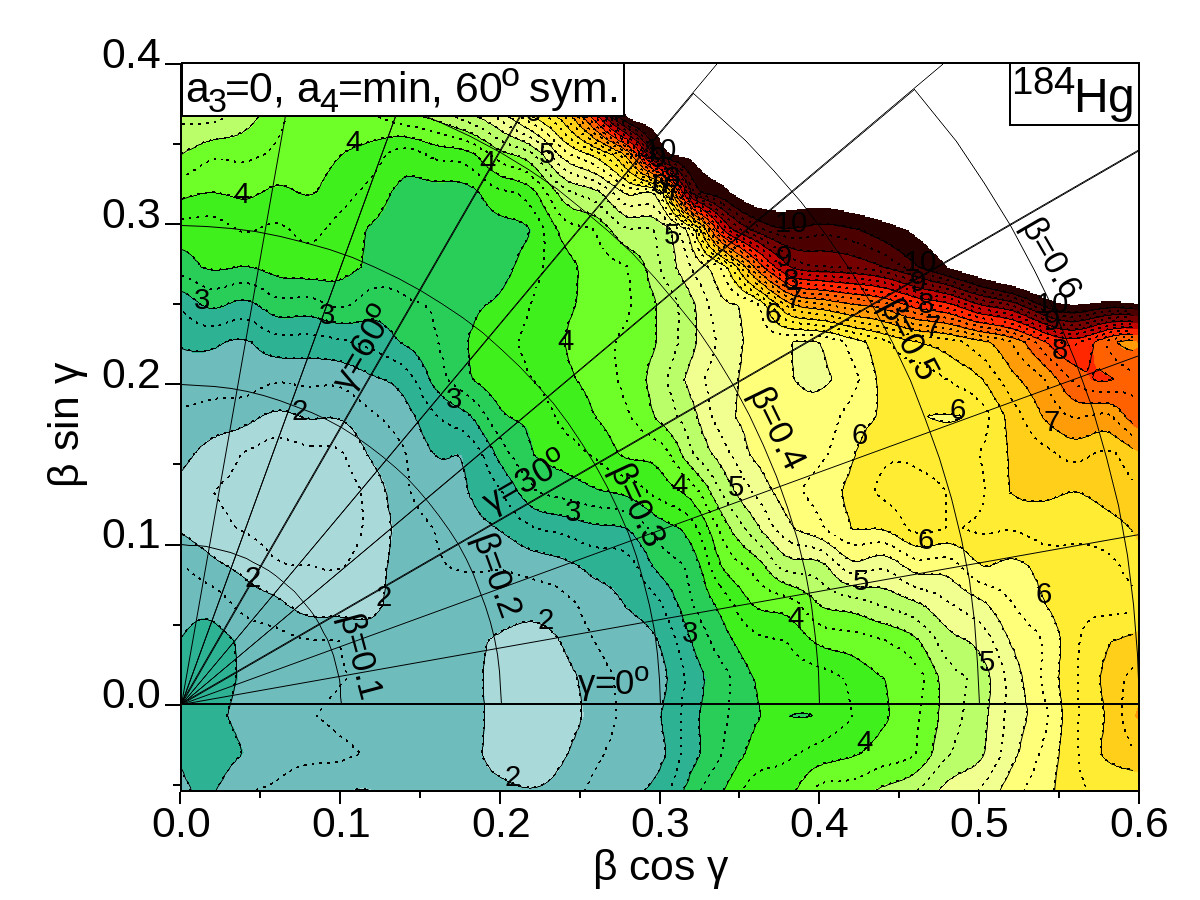} \\[0ex]
\includegraphics[width=\columnwidth]{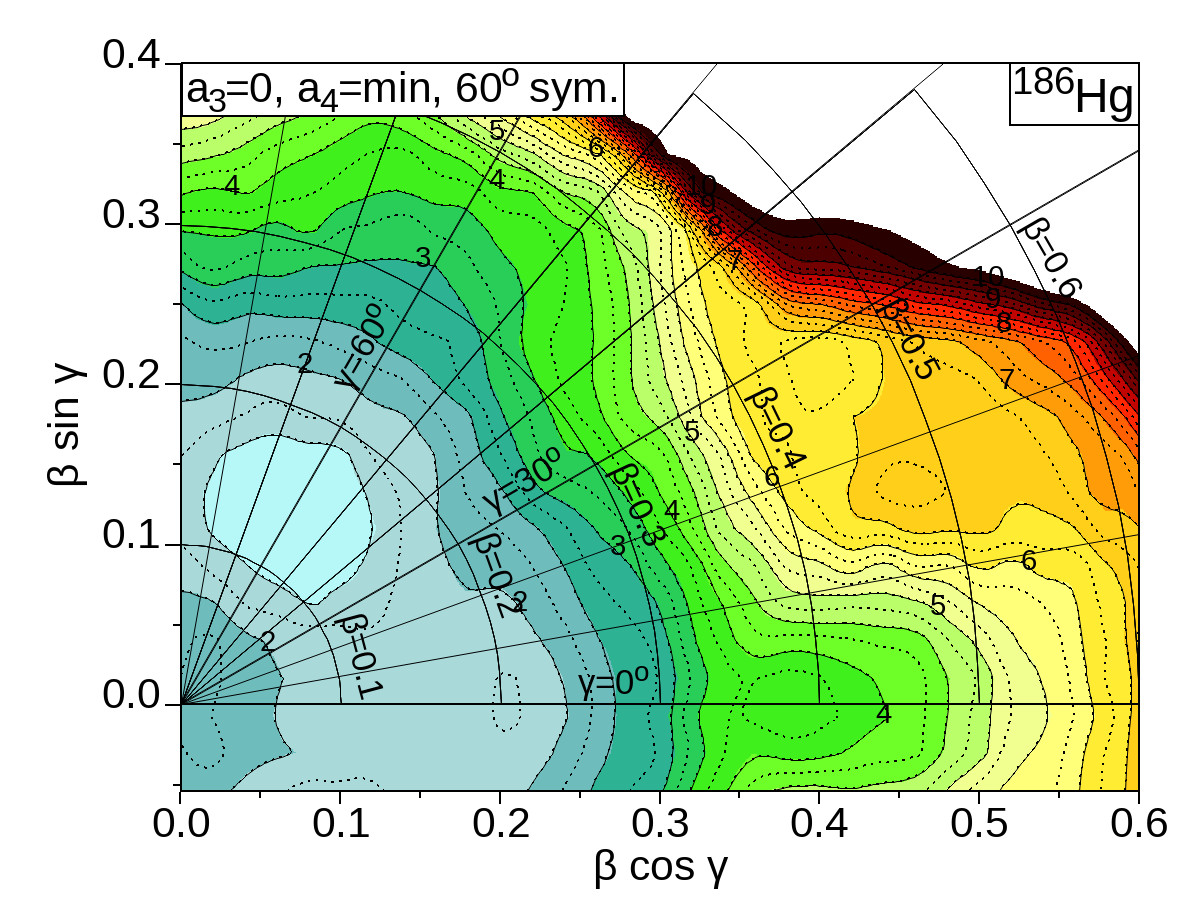} \\[0ex]
\includegraphics[width=\columnwidth]{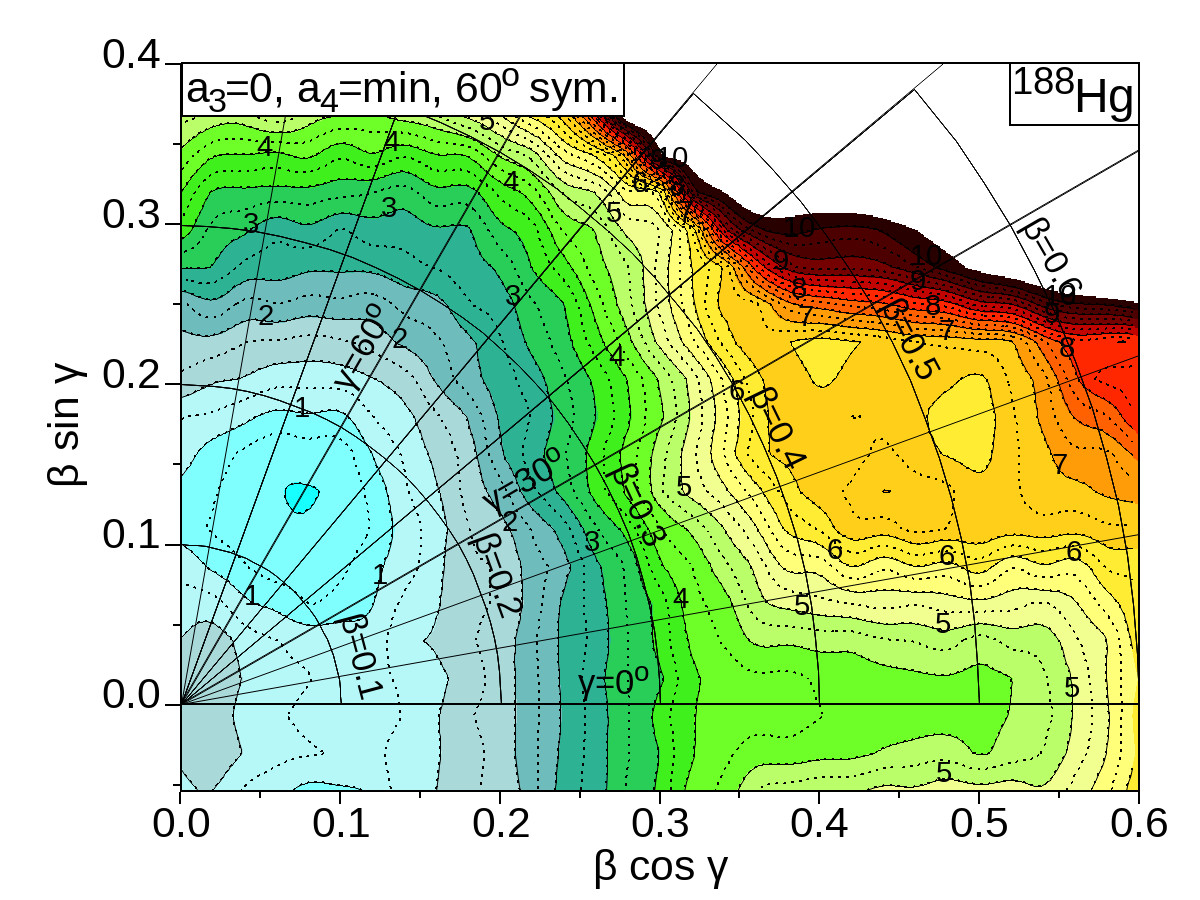} \\[0ex]
\caption{Symmetrized potential energy surfaces for Hg isotopes.}
\label{Hgmaps}
\end{figure}

\begin{figure}[!t]
\centering
\includegraphics[width=\columnwidth]{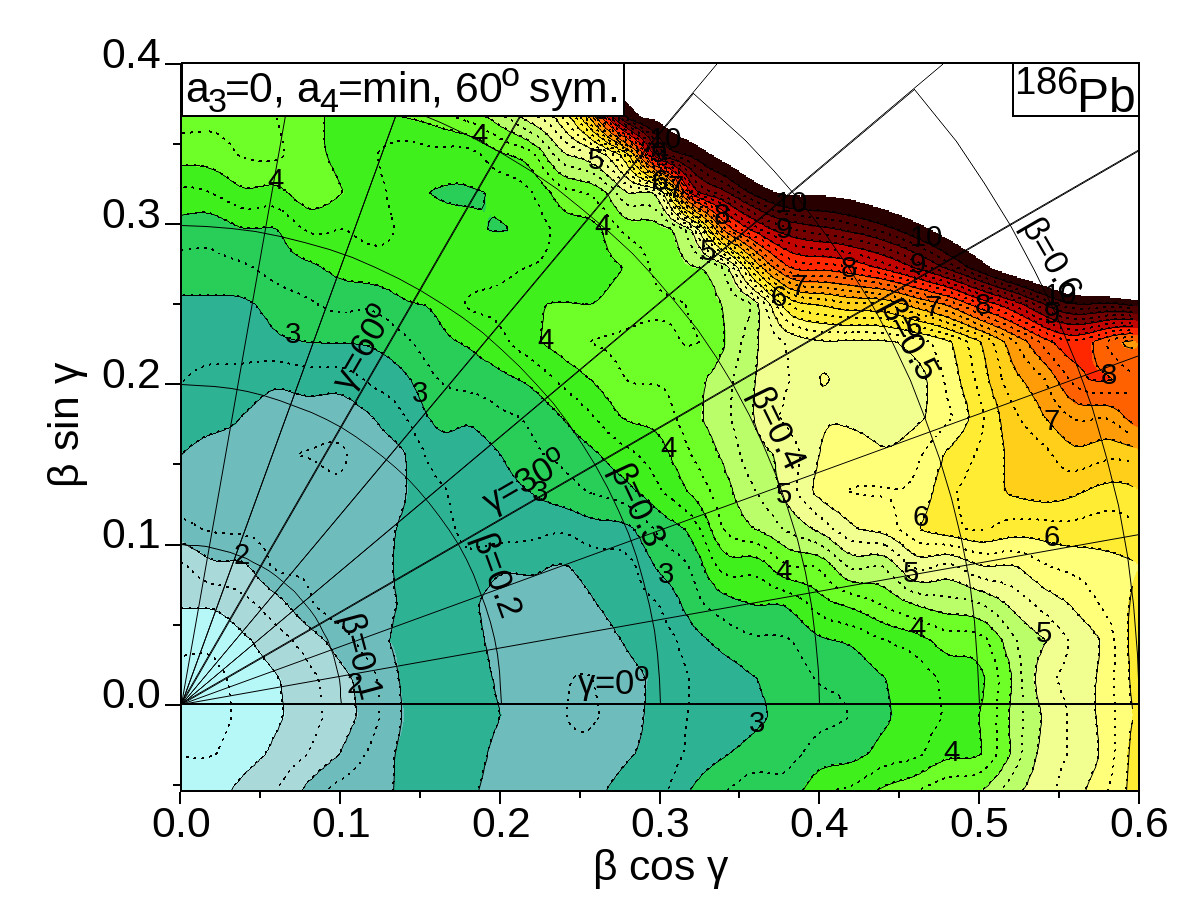} \\[0ex]
\includegraphics[width=\columnwidth]{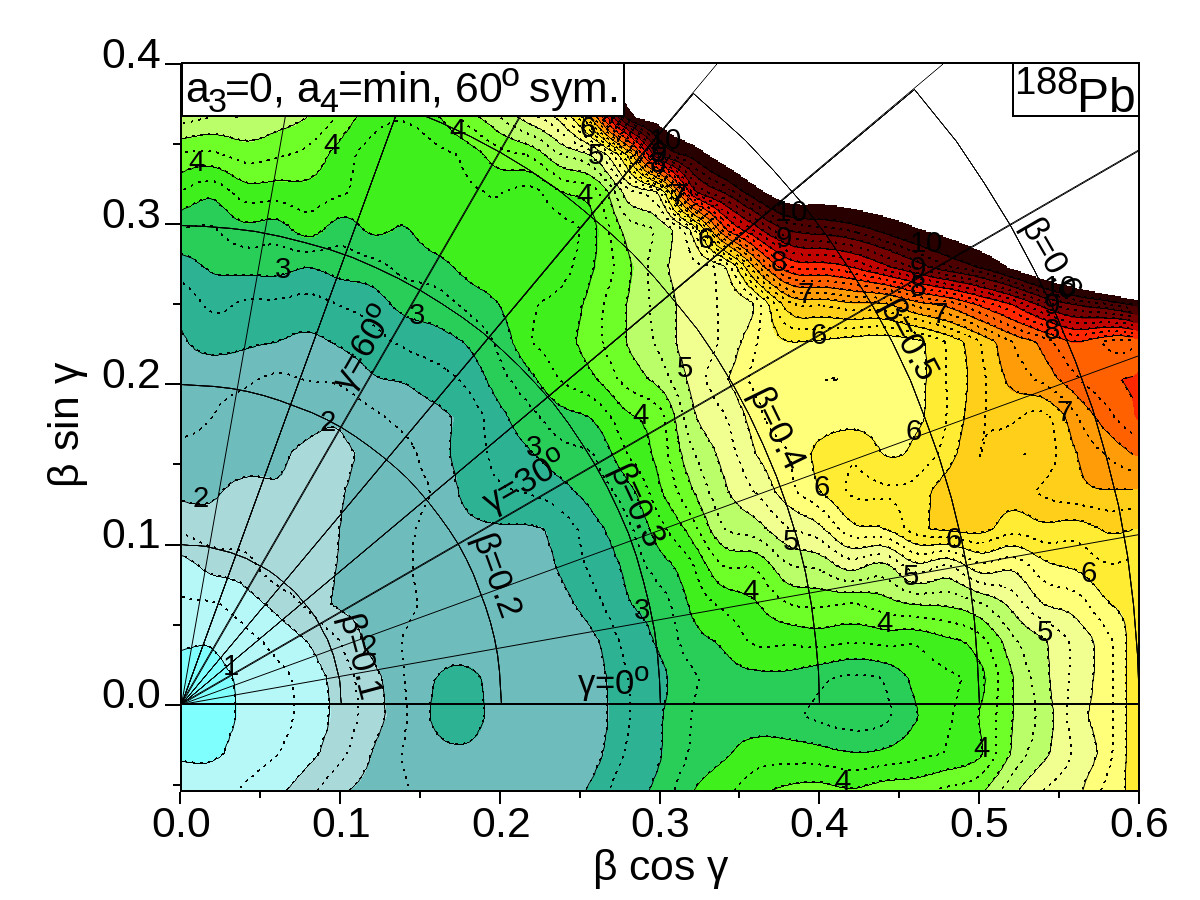} \\[0ex]
\includegraphics[width=\columnwidth]{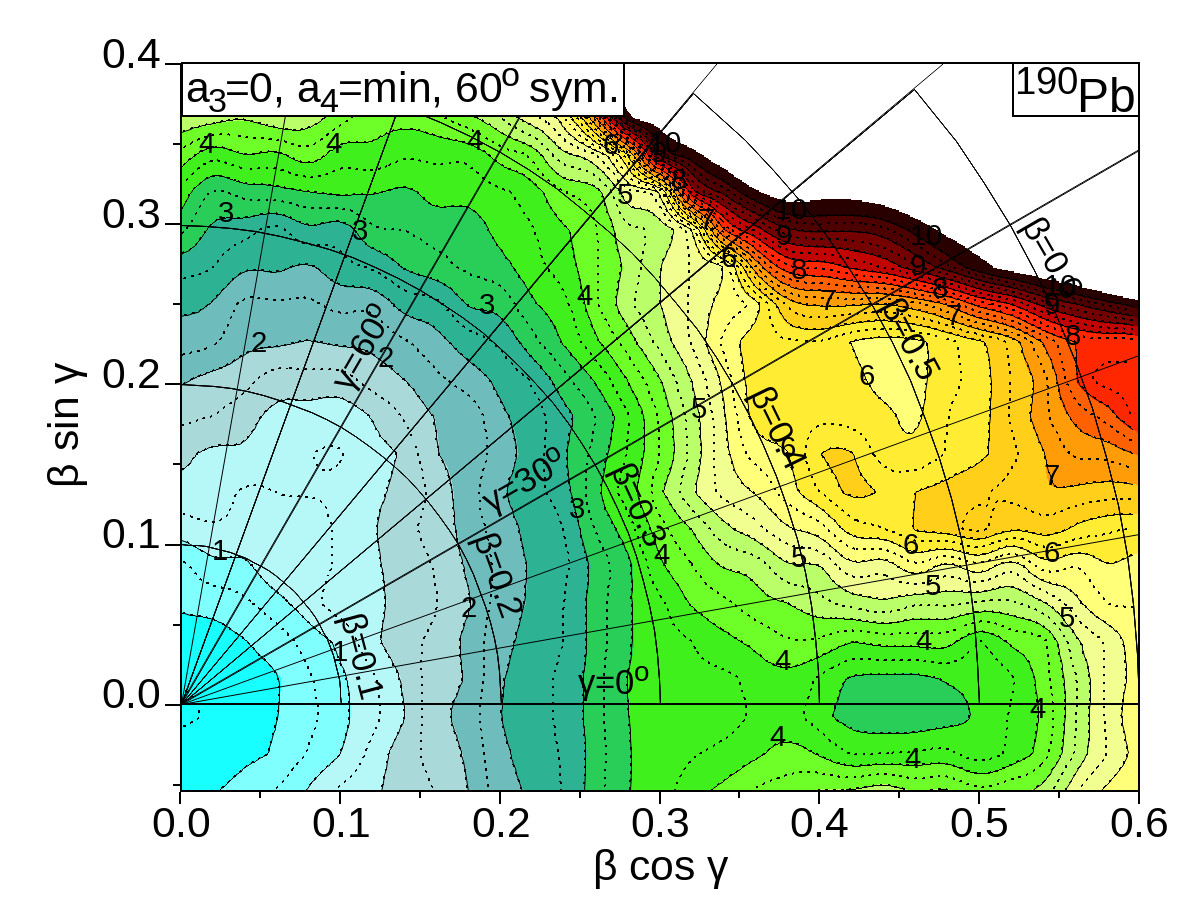} \\[0ex]
\caption{Symmetrized potential energy surfaces for Pb isotopes.}
\label{Pbmaps}
\end{figure}

We show in Figs. \ref{Ptmaps},$\;$\ref{Hgmaps} and \ref{Pbmaps} the symmetrized potential energy surfaces for the Pt, Hg, and Pb isotopes, obtained in the full 3-dimensional ($\{\beta,\gamma,a_4\}$) deformation space, but then minimized with respect to the neck parameter $a_4$ and projected on the traditional \cite{Boh52} $(\beta\cos\gamma, \beta\sin\gamma)$ plane. The importance of the left-right asymmetry parameter $a_3$ is found to be negligible for the considered nuclei at least at the small deformations $\beta < 0.6$ considered here. Consequently 
the SPESs shown in these figures have been obtained by imposing $a_3=0$. When studying very elongated shapes towards the end of our study (section IV), such a restriction is, however, abandoned.
%
%

For the platinum isotopes $^{182}$Pt up to $^{186}$Pt, a broad ground-state (g.s.) minimum is clearly visible at $\beta \approx 0.17$ and $\gamma=0$, and an oblate isomeric state ($\gamma \!=\! 60^\circ$) at slightly larger $\beta$ value. A barrier of approximately 0.5 MeV separates both minima. Please note that, for practical reasons, we represent in all the figures the full range $\gamma = 0-90^\circ$, where the symmetry of these figures with respect to the $\gamma = 60^\circ$-line should be noticed. When comparing the SPES of the three platinum isotopes shown here, one observes the emergence of some ``{\it valley}'' that extends between $\beta \approx 0.4$ and $0.5$ along the prolate axis. Such a valley will also be observed, even more pronounced, in the mercury and lead isotopes discussed below. In $^{186}$Pt, some bump emerges at $\{\beta \approx 0.43,\gamma \approx 20^\circ\}$ the appearance of which could already be expected when looking at the SPES of the $^{182}$Pt and $^{184}$Pt isotopes.

For the mercury isotope $^{184}$Hg, the picture is again about the same as for the platinum nuclei with a broad minimum clearly visible at $\beta \approx 0.15$, with the difference, however, that the ground state now evolves, with increasing mass number, to an oblate deformation ($\gamma = 60^{\circ}$). The prolate minimum at $\beta\approx 0.2$ is visible in $^{184}$Hg and marginally in the $^{186}$Hg isotope. A very small barrier separates the prolate and oblate minimum, barrier which disappears in $^{188}$Hg. In $^{184}$Hg a shallow shape isomer appears, in addition, at $\{\beta\approx 0.46,\gamma\approx 28^\circ\}$ around 6 MeV above the g.s. Similar non-axial isomeric states are also found in the two other mercury isotopes. In $^{188}$Hg an additional shape isomer appears at $\{\beta \approx 0.52,\gamma \approx 20^{\circ}\}$. Some broad prolate valley develops around 4 MeV above the g.s.\ in all three Hg isotopes at $\beta \approx $0.35, as already observed in the Pt nuclei, valley which stretches towards larger $\beta$ values as the neutron number increases.

The broad minimum along $\beta \approx 0.15$ observed for the platinum and mercury isotopes disappears in the lead nuclei from $^{186}$Pb up to $^{190}$Pb, in which the ground-state minimum appears at the spherical shape ($\beta=0$) and which becomes gradually deeper as the mass number increases and one approaches the doubly magic $^{208}$Pb. Please notice that there is a small oblate ($\gamma=60^\circ$) depression that appears in $^{186}$Pb at 
$\beta \approx 0.38$. The valley along the prolate axis that has already been observed at an elongation between $\beta=0.4$ and $\beta=0.5$ now leads in $^{188}$Pb and $^{190}$Pb nuclei to a clearly defined shape isomeric state at about 3 MeV above the ground state. 
In addition, a shallow oblate local minimum is found in the $^{190}$Pb isotope at $\beta=0.18$ and an energy of about 1.5 MeV.

\begin{figure}[!h]
\centering
\includegraphics[width=\columnwidth]{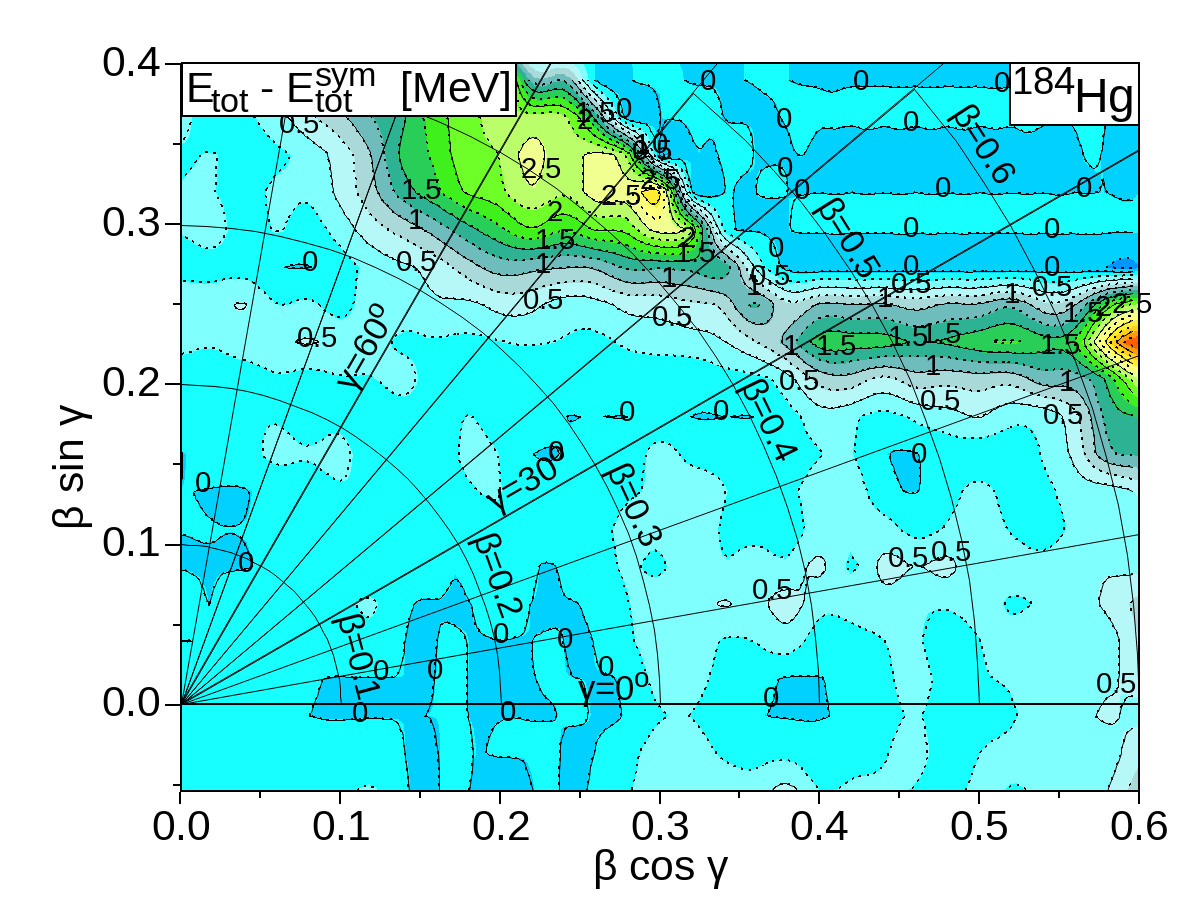} \\[0ex]
\caption{Energy gain obtained through the symmetrization procedure by taking into account the different orientations of the deformed nuclear shape in $^{184}$Hg.}
\label{184Hg_Egain}
\end{figure}

\begin{figure}[!h]
\centering
\includegraphics[width=\columnwidth]{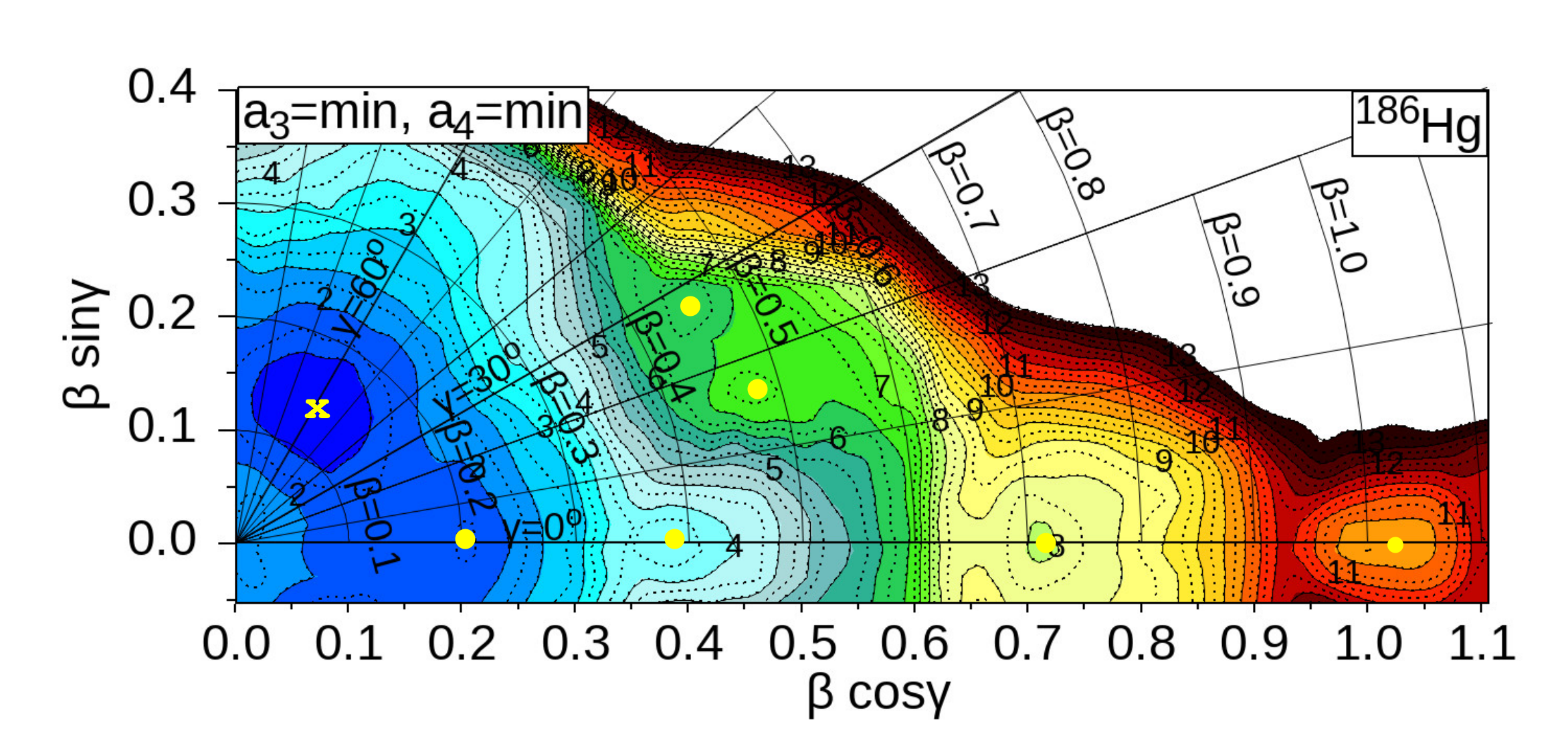} \\[0ex]
\includegraphics[width=\columnwidth]{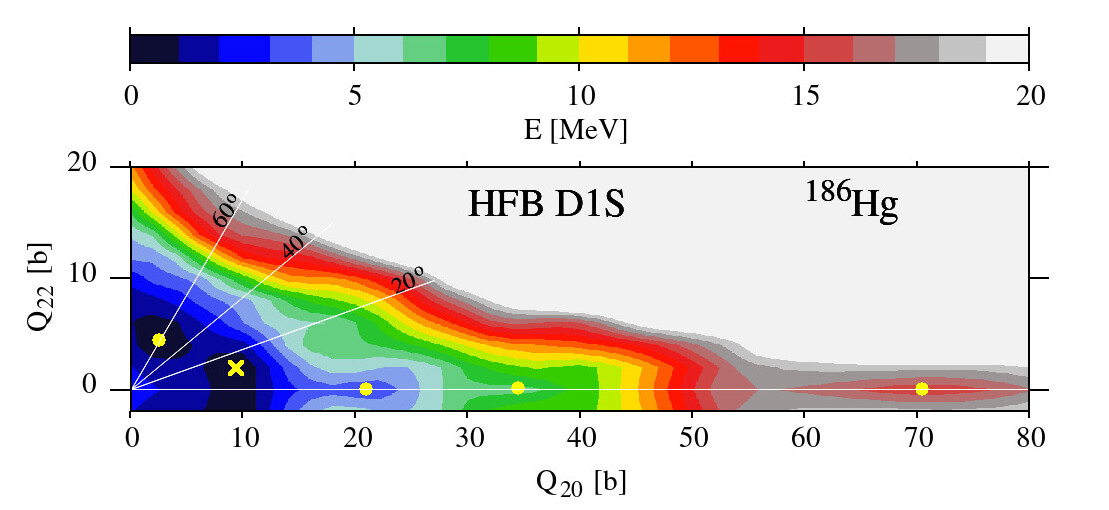} \\[0ex]
\caption{Potential energy surface of $^{186}$Hg evaluated within the macroscopic-microscopic approach (upper part) and, for comparison, with the HFB approach with the Gogny D1S force \cite{BGG84} (lower part), using constraints on the axial $Q_{20}$ and non-axial $Q_{22}$ mass quadrupole moments. Yellow dots mark local minima and the ground state is indicated by a cross.}
\label{186Hg_HFB}
\end{figure}
 
Let us emphasize at this point that whenever a nuclear shape goes beyond the simple tri-axial deformation as described by Bohr's $\{\beta,\gamma\}$ parametrization, i.e.\ whenever higher-order deformation terms in our Fourier-over-Spheroid shapes are taken into account, the different sectors $\{\beta,\gamma\}$, $\{\beta,\gamma +120^\circ\}$, and $\{\beta,\gamma +240^\circ\}$ (see Fig.\ \ref{Bohrbg}) are no longer going to be equivalent, as already pointed out in section II, and our symmetrization procedure that leads to what we have called the symmetrized potential energy surface (SPES) is going to take all its importance. To convince ourselves about the importance of what one could call the orientation degree of freedom of the nuclear shape, we are showing in Fig.\ \ref{184Hg_Egain} the energy gain obtained by minimizing the nuclear energy with respect to the three possible orientations of the deformed nuclear shape. One notices that this energy gain can indeed be quite substantial. As an illustrative example let us take the shape isomeric state found in the $^{184}$Hg isotope at $\{\beta \approx 0.46, \gamma \approx 28^\circ\}$ as shown in the top part of Fig.\ \ref{Hgmaps}. When looking now at Fig.\ \ref{184Hg_Egain} one has to conclude that at this deformation the energy gain due to our symmetrization procedure, is of the order of 1.5 MeV and this shape isomeric state would have been missed or certainly not properly described without taking the orientation degree of freedom into account.
Similar observations can be made practically for all the here described nuclei. The shape isomer found in $^{186}$Pb at $\beta \approx 0.36$ along the oblate axis $\gamma=60^\circ$ is only found because the energy gain due to the symmetrization procedure is as large as 1.5 MeV at that deformation and reaches 2.5 MeV for even larger $\beta$ values. 
It is therefore becoming obvious that for the analysis of possible nuclear shapes and the search for shape isomers, the present deformation study with the symmetrization procedure which takes the different possible orientation of the nuclear shape into account, is much more precise as compared to the approach of Ref.\ \cite{PND20} where all of the above shape isomers would have been missed.

The question now arises as to how reliable our estimates really are. Trying to give an 
answer, we have performed for the isotope $^{186}$Hg a self-consistent constrained HFB calculation with the Gogny D1S force \cite{BGG84} using as a constraint the axial $Q_{20}$ and non-axial $Q_{22}$ mass quadrupole moments. This method allows to exhibit the important role of triaxial deformation around the ground state shape of nuclei from the Pt region \cite{RRS09,RSRG10}.
The corresponding PES for $^{186}$Hg is shown in the lower part of Fig.\ \ref{186Hg_HFB}. In this HFB calculation, the ground state is found for an essentially prolate shape with a slight triaxiality of about $\gamma=12^{\circ}$, about 200 keV lower than the oblate minimum. In the macroscopic-microscopic approach (upper part of Fig.\ \ref{186Hg_HFB}), the ground state is found, on the contrary, to be oblate deformed about 300 keV lower than the prolate minimum. In both calculations, several prolate local minima are found, in the HFB calculation at $Q_{20} \approx 21.0$, 34.4 and  70.5 b, corresponding respectively to a value of $\beta=0.56$, 0,79 and 1,13 in the macroscopic-microscopic calculation, when using the rough estimate 
\begin{equation}
    \beta=\frac{4 \pi}{5} \frac{\sqrt{Q_{20}^2+Q_{22}^2}}{A r_{\mathrm{rms}}^2}\;.
\end{equation}

The corresponding energies are found, respectively, at about $E=3.0$,  6.5  and 15.3 MeV above the ground state. There is a  clear correspondence with the shape of the potential energy surface of the LSD model where similar minima are found, respectively, at $\beta=0.38$, 0.71 and 1.01 with energies $E=2.3$,  6.6 and 8.9 MeV above the ground-state. Despite some  quantitative differences, one can notice an equivalence in the structure of this nucleus in both descriptions.


\section{Higher deformations effect}

\begin{figure}[!h]
\centering
\includegraphics[width=\columnwidth]{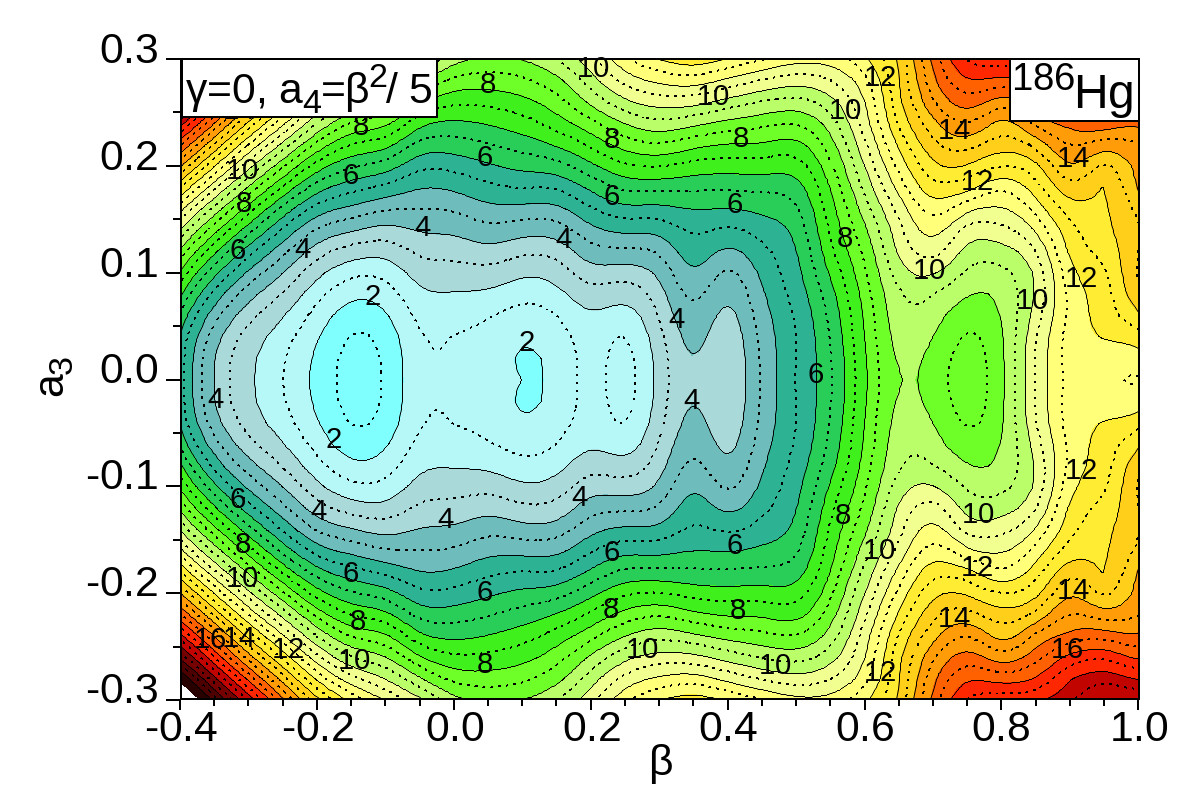} \\
\includegraphics[width=\columnwidth]{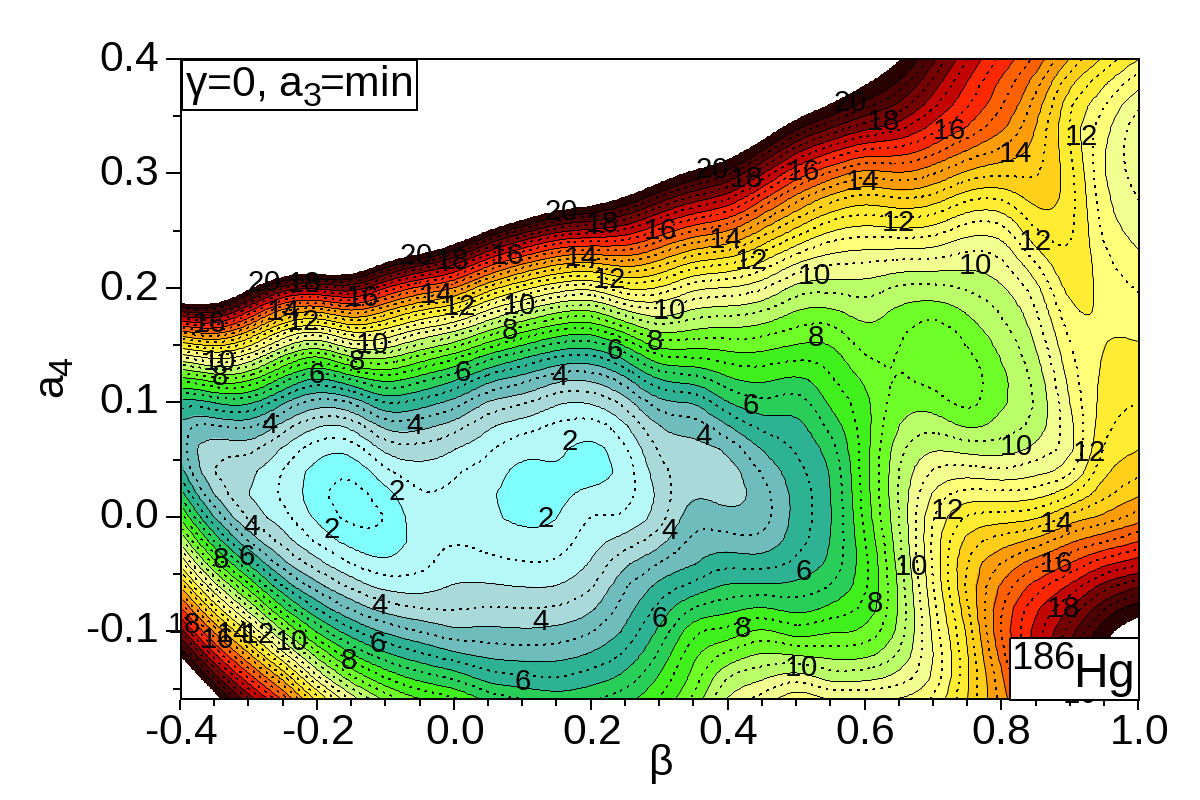} \\[0ex]
\caption{Potential energy surface of $^{186}$Hg for the axially symmetric case ($\gamma=0$) as function of $\beta$ and $a_3$ (upper map) and of $\beta$ and $a_4$ (lower map). }
\label{186Hg_eb34m}
\end{figure}

Several triaxial local minima have been found in the above studied nuclei, mainly at an elongation of $\beta \approx 0.5$ and a rather small non-axiality of $\gamma \approx 25^{\circ}$. The above presented landscapes (Figs.\ 3 to 5) have been obtained by imposing left-right symmetric shapes ($a_3=0$) and by minimizing the total nuclear energy with respect to what one could call the neck parameter $a_4$, except for the calculations of the SPES of $^{186}$Hg (Fig.\ \ref{186Hg_HFB}), where a minimization is carried out with respect to both these parameters. To really study the importance of possible octupole deformations we have studied the SPES of $^{186}$Hg by plotting the nuclear energy as a function of ($\beta,\,a_3$) now imposing axial symmetry ($\gamma=0$) and assuming $a_4=\beta^2/5$ which corresponds approximately to the bottom of the valley in the ($\beta,\,a_4$) plane (see upper part of Fig.\ \ref{186Hg_eb34m}). One thus finds that the minimal-energy deformation is obtained for left-right symmetric shapes ($a_3=0$).
%
%
Considering only this figure, one would be tempted to conclude at the possible presence of two minimal-energy configurations, corresponding one to a prolate ($\beta \approx 0.20$) and the other to an oblate shape ($\beta \approx -0.15$) 
separated by a barrier of about 0.5 MeV. However, 
%
%
%
Fig.\,\ref{Hgmaps} does not support the existence of a prolate minimum since the barrier between both minima is negligibly small.
The tendency to form a rather elongated prolate shape around $\beta \approx 0.725$, clearly visible in Fig.\,\ref{186Hg_eb34m}, seems, on the contrary, to be confirmed by a minimization calculation with respect to $a_3$ and $a_4$. 
%
Another minimum similar to the one visible in Fig.~\ref{186Hg_HFB} at $Q_{20}$=70 b is formed at $\beta > 1.0$ in Fig.~\ref{186Hg_eb34m}.


\section{Summary and conclusion}

In our study of shape isomers in even-even nuclei of the 
isotopic chains of Pt, Hg, and Pb around $^{186}$Hg, a certain number of general remarks should be made. First of all, we have found that the left-right asymmetry degree of freedom does not play any important role in any of the here-considered nuclei, and this not only around the ground state but throughout the whole range of deformations, as long as the elongation does not become too large ($\beta < 0.6$).
Let us mention at this point, however, that this conclusion does not necessarily hold when going to very large elongations, as it is well known that the left-right asymmetry becomes important for elongations encountered in the fission process. On the contrary, the minimization of the SPESs with respect to the neck deformation parameter $a_4$, 
essentially the hexadecupole degree of freedom, 
makes the ground state minima in Pt and Hg isotopes deeper. It is, therefore, important for the study of shape isomers. 

A prolate-oblate shape coexistence has been shown to be  possible in the considered Pt and Hg isotopes, 
where the ground-state deformation is found on the prolate side in the platinium isotopes with a tendency, with increasing mass number, towards triaxial shapes (increasing $\gamma$ value). In all three platinium isotopes, a clearly identified oblate shape isomer is found
around $\beta \approx 0.18$.

In the neighboring mercury isotopes, the ground-state deformation is found on the oblate side, with a prolate shape isomer. With increasing mass numbers, the ground state becomes successively deeper while the prolate shape isomer gets shallower. In all three mercury isotopes, triaxial shape isomeric states are found around $\gamma \approx 20^{\circ}$ at an elongation of $\beta \approx 0.5$.

In the lead isotopes, the ground-state deformation has clearly shifted towards a spherical shape with an oblate shape-isomeric state around $\beta \approx 0.2$ and a prolate shape isomer that develops with increasing mass number at an elongation of $\beta \approx 0.45$. A triaxial shape isomer is found in all three lead isotopes at 
$\{\beta \approx 0.2, \gamma \approx 30^{\circ}\}$.

It has also been found that for a correct description of shape isomeric states 
in terms of the very widely used Bohr shape parameters $\{\beta,\gamma\}$
it is absolutely essential to include all possible orientations in space of the deformed shape. That means that, as soon as higher order deformations are taken into account, there is no longer any redundance in the three 120$^{\circ}$ $\gamma$-sectors in the Bohr $\{\beta,\gamma\}$ shape parametrization and it turns out to be of capital importance to include a symmetrization procedure in the calculations which takes care of this orientation degree of freedom of the nuclear shape, thus leading to what has been called a ``{\it symmetrized potential energy surface}'' (SPES). 
It has also been shown that the potential energy surfaces obtained in our present model are, indeed, close to the ones obtained in the HFB theory with the Gogny effective interaction. We have, in particular,  predicted in both models the existence of three prolate shape isomers in $^{186}$Hg.

Future investigations of other nuclear isotopic chains in the $Z \approx 82$ and other very different mass regions should be considered to probe, from a comparison with the experimental data, the predictive power of our theoretical approach.\\
~\\

{\bf Acknowledgements}\\
Our research was supported by the Polish-French collaboration agreement COPIN-IN2P3: project No. 08-131 and 15-149.
This research was also funded in part by the National Science Centre, Poland under research project No.\ 2023/49/B/ST2/01294. 


\end{document}